\documentclass[conference]{IEEEtran}
\IEEEoverridecommandlockouts
\usepackage{url}

\usepackage{listings}
\usepackage{color}
\usepackage{float}

\usepackage{capt-of}

\definecolor{codegreen}{rgb}{0,0.6,0}
\definecolor{codegray}{rgb}{0.5,0.5,0.5}
\definecolor{codepurple}{rgb}{0.58,0,0.82}
\definecolor{backcolour}{rgb}{0.95,0.95,0.92}
\definecolor{lightblue}{rgb}{0.67,0.84,0.90}

\lstdefinestyle{mystyle}{
	backgroundcolor=\color{backcolour},   
	commentstyle=\color{codegreen},
	keywordstyle=\color{magenta},
	numberstyle=\tiny\color{codegray},
	stringstyle=\color{codepurple},
	basicstyle=\scriptsize,
	breakatwhitespace=false,
	escapeinside={\%*}{*)},
	breaklines=true,
	captionpos=t,
	keepspaces=true,
	numbers=left,
	numbersep=5pt,
	showspaces=false,
	showstringspaces=false,
	showtabs=false,
	tabsize=2,
	mathescape=true
}

\usepackage{todonotes}

\usepackage{graphics}

\usepackage{graphicx}
\usepackage{caption}
\usepackage{subcaption}

\usepackage{tikz}
\newcommand*\circled[1]{\tikz[baseline=(char.base)]{
		\node[shape=circle,draw,inner sep=0.5pt] (char) {#1};}}

\usepackage{multirow}
\usepackage{diagbox}

\let\labelindent\relax
\usepackage{enumitem}

\usepackage{bbding}

\usepackage{cite}

\ifCLASSINFOpdf
\else
\fi
\begin{document}
%
\title{A Primer on RecoNIC: \underline{R}DMA-\underline{e}nabled \underline{C}ompute \underline{O}ffloading on Smart\underline{NIC}}

\setlength{\parskip}{0pt}

\author{
	\IEEEauthorblockN{
		Guanwen Zhong\IEEEauthorrefmark{1} \enspace
		Aditya Kolekar\IEEEauthorrefmark{1}\IEEEauthorrefmark{3} \enspace
		Burin Amornpaisannon\IEEEauthorrefmark{1}\IEEEauthorrefmark{3} \enspace
		Inho Choi\IEEEauthorrefmark{1}\IEEEauthorrefmark{3} \enspace
		Haris Javaid\IEEEauthorrefmark{1} \enspace
		Mario Baldi\IEEEauthorrefmark{2}
	}
	\IEEEauthorblockA{
		\IEEEauthorrefmark{1}AMD, Singapore \quad 
		\IEEEauthorrefmark{2}AMD, USA
		\IEEEcompsocitemizethanks{\IEEEcompsocthanksitem\IEEEauthorrefmark{3}Work done during internship at AMD/Xilinx. Aditya is also affiliated with NTU, Singapore, and Burin and Inho are now at NUS, Singapore.}
	}
	\IEEEauthorblockA{henry.zhong@amd.com}
}


%


\maketitle


%
\IEEEpeerreviewmaketitle

\begin{abstract}
Today's data centers consist of thousands of network-connected hosts, each with CPUs and accelerators such as GPUs and FPGAs. These hosts also contain network interface cards (NICs), operating at speeds of 100Gb/s or higher, that are used to communicate with each other. We propose RecoNIC, an FPGA-based RDMA-enabled SmartNIC platform that is designed for compute acceleration while minimizing the overhead associated with data copies (in CPU-centric accelerator systems) by bringing network data as close to computation as possible. Since RDMA is the defacto transport-layer protocol for improved communication in data center workloads, RecoNIC includes an RDMA offload engine for high throughput and low latency data transfers. Developers have the flexibility to design their accelerators using RTL, HLS or Vitis Networking P4 within the RecoNIC's programmable compute blocks. These compute blocks can access host memory as well as memory in remote peers through the RDMA offload engine. Furthermore, the RDMA offload engine is shared by both the host and compute blocks, which makes RecoNIC a very flexible platform. Lastly, we have open-sourced RecoNIC for the research community to enable experimentation with RDMA-based applications and use-cases.
\end{abstract}

\section{Introduction}
\label{sec:intro}

\begin{figure}[t!]
	\centering
	\includegraphics[width=0.9\columnwidth]{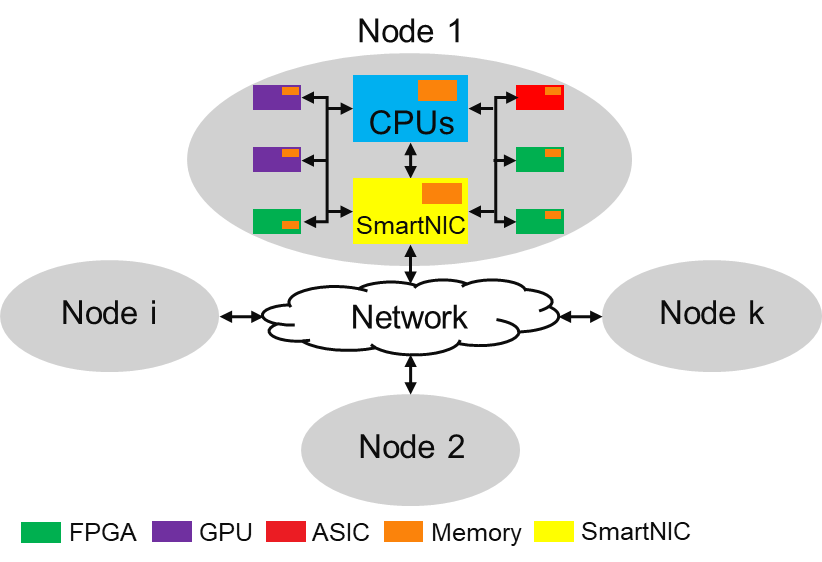}
	\caption{An example of data center architectures}
	\label{datacenter_arch}
\end{figure}

To meet the explosive growth of data and workloads/applications, today's data centers comprise a sea of network-connected hosts, each with multi-core CPUs and accelerators in the form of ASICs, FPGAs, and/or GPUs. These hosts communicate with each other via network interface cards (NICs). Over just a few years, data center networking throughput has surged by over 100 times, advancing from 1GbE to 100GbE and beyond. This remarkable growth far surpasses the evolution of CPUs in terms of computational capacity. The traditional reliance on CPUs for network processing and data transmission at such high bandwidths results in heightened CPU utilization, leading to longer latency, reduced throughput, and increased costs for delivering data center services to customers~\cite{azure_networking}. This has led to the emergence of SmartNIC technologies~\cite{azure_networking},\cite{nv_bluefield} by offloading network functionalities from host CPUs onto SmartNICs. In this data center architecture, SmartNICs play a crucial role and act as intermediate hubs, as depicted in Fig.~\ref{datacenter_arch}. All the network traffic passes through a SmartNIC to various compute elements, such as CPUs, GPUs, FPGAs, and ASICs, as well as external network-connected peers. Various SmartNIC architectures exist, including ASIC-based solutions (e.g., NVIDIA BlueField~\cite{nv_bluefield} powered with RDMA offloading and ARM processors), P4-programmable-based designs (e.g., AMD Pensando~\cite{pensando}), and FPGA-based SmartNICs (e.g., AMD Alveo SN1000~\cite{sn1000} and Microsoft Azure SmartNIC~\cite{azure_networking}). Despite the steeper learning curve associated with FPGA-based SmartNICs in comparison to ASIC and P4-programmable solutions, Microsoft Azure data centers~\cite{azure_networking} have demonstrated the mass deployment of FPGA-based SmartNICs, achieving high performance and efficiency that are not feasible with CPUs and offering programmability far beyond what an ASIC can provide, all at a reasonable cost. In this work, we primarily focus on FPGA-based SmartNIC solutions.

The ever increasing demands of data center workloads, including machine learning training/inference and high performance computing applications, necessitate high-speed networking with high throughput, low latency, and minimal CPU overhead. This demand has driven the emergence of remote direct memory access (RDMA) as the de facto standard for high-speed data center networking. RoCEv2 (RDMA over Converged Ethernet Version 2) stands out as a popular RDMA protocol that is well-supported and offloaded in ASIC-based SmartNICs like NVIDIA BlueField~\cite{nv_bluefield}. However, existing FPGA-based SmartNICs~\cite{azure_networking},\cite{open_nic},\cite{corundum},\cite{qep_driver} exhibit limited support for RDMA offloading.

\begin{table*}[t!]
	\caption{State-of-the-art FPGA-based Networking and SmartNIC Platforms.}
	\label{tab:networking_platforms}
	\begin{tabular}{lccccccc}
		\hline
		\multicolumn{1}{|l|}{\multirow{2}{*}{Reference}} & \multicolumn{1}{c|}{\multirow{2}{*}{Platform Type}} & \multicolumn{1}{c|}{\multirow{2}{*}{Software Network Stack}} & \multicolumn{1}{c|}{\multirow{2}{*}{Transport-layer Offloading}} & \multicolumn{1}{c|}{\multirow{2}{*}{\begin{tabular}[c]{@{}c@{}}Supported RDMA \\ Operation\end{tabular}}}                                   & \multicolumn{2}{c|}{Compute Offloading}                         & \multicolumn{1}{c|}{\multirow{2}{*}{Open-source}} \\ \cline{6-7}
		\multicolumn{1}{|l|}{}                           & \multicolumn{1}{c|}{}                               & \multicolumn{1}{c|}{}                                        & \multicolumn{1}{c|}{}                                            & \multicolumn{1}{c|}{}                                                                                                                       & \multicolumn{1}{c|}{Lookaside} & \multicolumn{1}{c|}{Streaming} & \multicolumn{1}{c|}{}                             \\ \hline
		\multicolumn{1}{|l|}{AlveoLink~\cite{alveolink}}                  & \multicolumn{1}{c|}{FPGA Networking}                & \multicolumn{1}{c|}{\XSolidBrush}                                      & \multicolumn{1}{c|}{HLS-based RDMA}                              & \multicolumn{1}{c|}{Not clear}                                                                                                              & \multicolumn{1}{c|}{\XSolidBrush}        & \multicolumn{1}{c|}{\Checkmark}       & \multicolumn{1}{c|}{\Checkmark}                          \\ \hline
		\multicolumn{1}{|l|}{XUP VNx~\cite{VNx}}                    & \multicolumn{1}{c|}{FPGA Networking}                & \multicolumn{1}{c|}{\XSolidBrush}                                      & \multicolumn{1}{c|}{UDP}                                         & \multicolumn{1}{c|}{\XSolidBrush}                                                                                                                     & \multicolumn{1}{c|}{\XSolidBrush}        & \multicolumn{1}{c|}{\Checkmark}       & \multicolumn{1}{c|}{\Checkmark}                          \\ \hline
		\multicolumn{1}{|l|}{Coyote~\cite{coyote}}                     & \multicolumn{1}{c|}{FPGA Networking}                & \multicolumn{1}{c|}{\XSolidBrush}                                      & \multicolumn{1}{c|}{TCP and HLS-based RDMA}                      & \multicolumn{1}{c|}{Read/Write/Send}                                                                                                        & \multicolumn{1}{c|}{\Checkmark}       & \multicolumn{1}{c|}{\Checkmark}       & \multicolumn{1}{c|}{\Checkmark}                          \\ \hline
		\multicolumn{1}{|l|}{FpgaNIC~\cite{fpganic}}                    & \multicolumn{1}{c|}{FPGA Networking}                & \multicolumn{1}{c|}{\XSolidBrush}                                      & \multicolumn{1}{c|}{TCP}                                         & \multicolumn{1}{c|}{\XSolidBrush}                                                                                                                     & \multicolumn{1}{c|}{\Checkmark}       & \multicolumn{1}{c|}{\Checkmark}       & \multicolumn{1}{c|}{\Checkmark}                          \\ \hline
		\multicolumn{1}{|l|}{COPA~\cite{copa}}                       & \multicolumn{1}{c|}{FPGA Networking}                & \multicolumn{1}{c|}{\XSolidBrush}                                      & \multicolumn{1}{c|}{Non-RoCEv2 RDMA}                             & \multicolumn{1}{c|}{Put/Get/Send}                                                                                                           & \multicolumn{1}{c|}{\Checkmark}       & \multicolumn{1}{c|}{\Checkmark}       & \multicolumn{1}{c|}{\XSolidBrush}                           \\ \hline
		\multicolumn{1}{|l|}{Azure SmartNIC~\cite{azure_networking}}                        & \multicolumn{1}{c|}{SmartNIC}                       & \multicolumn{1}{c|}{\Checkmark}                                     & \multicolumn{1}{c|}{LTL Engine*}                   & \multicolumn{1}{c|}{\XSolidBrush}                                                                                                                     & \multicolumn{1}{c|}{\Checkmark}       & \multicolumn{1}{c|}{\Checkmark}       & \multicolumn{1}{c|}{\XSolidBrush}                           \\ \hline
		\multicolumn{1}{|l|}{QEP~\cite{qep_driver}}                        & \multicolumn{1}{c|}{SmartNIC}                       & \multicolumn{1}{c|}{\Checkmark}                                     & \multicolumn{1}{c|}{TCP/UDP Checksum and RSS\textsuperscript{$\dagger$}}                   & \multicolumn{1}{c|}{\XSolidBrush}                                                                                                                     & \multicolumn{1}{c|}{\Checkmark}       & \multicolumn{1}{c|}{\Checkmark}       & \multicolumn{1}{c|}{\XSolidBrush}                           \\ \hline
		\multicolumn{1}{|l|}{OpenNIC~\cite{open_nic}}                    & \multicolumn{1}{c|}{SmartNIC}                       & \multicolumn{1}{c|}{\Checkmark}                                     & \multicolumn{1}{c|}{\XSolidBrush}                                          & \multicolumn{1}{c|}{\XSolidBrush}                                                                                                                     & \multicolumn{1}{c|}{\XSolidBrush}        & \multicolumn{1}{c|}{\Checkmark}       & \multicolumn{1}{c|}{\Checkmark}                          \\ \hline
		\multicolumn{1}{|l|}{Corundum~\cite{corundum}}                   & \multicolumn{1}{c|}{SmartNIC}                       & \multicolumn{1}{c|}{\Checkmark}                                     & \multicolumn{1}{c|}{\XSolidBrush}                                          & \multicolumn{1}{c|}{\XSolidBrush}                                                                                                                     & \multicolumn{1}{c|}{\XSolidBrush}        & \multicolumn{1}{c|}{\Checkmark}       & \multicolumn{1}{c|}{\Checkmark}                          \\ \hline \hline
		\multicolumn{1}{|l|}{RecoNIC}                    & \multicolumn{1}{c|}{SmartNIC}                       & \multicolumn{1}{c|}{\Checkmark}                                     & \multicolumn{1}{c|}{RDMA}                                        & \multicolumn{1}{c|}{\begin{tabular}[c]{@{}c@{}}Read/Write/Send/\\ Write with IMMDT/\\ Send with IMMDT/\\ Send with invalidate\end{tabular}} & \multicolumn{1}{c|}{\Checkmark}       & \multicolumn{1}{c|}{\Checkmark}       & \multicolumn{1}{c|}{\Checkmark}                          \\ \hline
		\multicolumn{8}{l}{*Lightweight transport-layer engine \quad $\dagger$Receive Side Scaling support }                                                                                                                                                                                                                                
	\end{tabular}
\end{table*}

We introduce RecoNIC, an open-source\footnote{\label{reconic_repo_footnote}Available at https://github.com/Xilinx/RecoNIC} RDMA-enabled SmartNIC platform with compute acceleration, designed to enable network-attached compute acceleration and minimize the overhead associated with data copies. Developers have the flexibility to design their accelerators using RTL, HLS, or Vitis Networking P4 within RecoNIC's programmable compute blocks. This enables the processing of network data without multiple data copies' that are typically the characteristic of traditional CPU-centric solutions. With an FPGA-based SmartNIC, the traditional CPU-centric solution can be shifted to an FPGA-centric solution by storing network data in the SmartNIC's device memory, significantly reducing data copies over PCIe and bringing data as close to computation as possible. The logic executed within the compute blocks can access both host and device memory in remote peers via RDMA. Our contributions can be summarized as follows:

\begin{itemize}
	\item RecoNIC, an open-source\textsuperscript{1} FPGA-based RDMA-enabled SmartNIC platform.
	\item RecoNIC supports both RDMA and non-RDMA traffic, as well as streaming and lookaside compute acceleration. The compute blocks allow users to develop accelerators via Vitis Networking P4, HLS, and RTL.
	\item The RDMA offloading engine is shared by both host CPUs (over PCIe) and FPGA compute blocks, which makes the platform very flexible. RDMA Queue Pairs (QPs) can be allocated either on host or device memory.
\end{itemize}

\section{Related Work}
FPGA-based networking solutions can be categorized into two groups: FPGA-based networking platforms and FPGA-based SmartNIC platforms. FPGA-based networking platforms can work as standalone nodes and are equipped with transport-layer offloading engines and a media access control (MAC) component, enabling adaptive accelerators to communicate with each other via the network. FPGA-based SmartNICs act as NICs to provide standard networking functionality for the host CPUs in addition to some advanced networking features. The main differences between the two categories are as follows:
\begin{itemize}
	\item FPGA-based SmartNIC platforms can replace traditional network interface cards (NICs) and offer standard NIC functionalities, such handling of various protocols through a software network stack, traffic management, and virtual switch support.
	\item FPGA-based SmartNIC platforms can support additional NIC functionalities like network measurement and telemetry for various protocols with or without offloading engines.
	\item FPGA-based SmartNIC platforms enable processing of layer-4 and above protocols that are not offloaded as transport-layer protocols in hardware.
\end{itemize}

An overview of the existing state-of-the-art FPGA-based networking and SmartNIC platforms is presented in Table~\ref{tab:networking_platforms}. FPGA-based networking platforms~\cite{alveolink},\cite{VNx},\cite{copa},\cite{fpganic},\cite{coyote} are equipped with transport-layer offloading engines and establish communication with other FPGA boards connected through these engines. However, due to the absence of a software network stack, these platforms cannot communicate with other devices (either FPGA boards or traditional NICs) over layer-4 protocols that are not supported by their transport-layer offloading engines. In contrast, FPGA-based SmartNIC platforms~\cite{azure_networking},\cite{open_nic},\cite{corundum},\cite{qep_driver} employ a software network stack to communicate with other devices, regardless of whether they are FPGA boards or traditional NICs. Nevertheless, the majority of FPGA-based SmartNIC platforms~\cite{open_nic},\cite{corundum},\cite{qep_driver}, except for Microsoft Azure SmartNIC~\cite{azure_networking}, which is powered by a lightweight transport-layer offloading engine, heavily rely on host CPUs for processing network data. This reliance results in underutilization of network bandwidth and high latency, especially in high-speed networks ($>=$ 100Gb/s).

RDMA serves as the de facto standard for high-speed data communication in machine learning and high performance computing applications. Notably, the state-of-the-art FPGA-based SmartNIC platforms~\cite{azure_networking},\cite{open_nic},\cite{corundum},\cite{qep_driver} do not support RDMA. Among FPGA-based networking platforms, three works~\cite{alveolink},\cite{copa},\cite{coyote} include RDMA offloading engines. COPA [15] incorporates a non-RoCEv2 RDMA offloading engine, while AlveoLink~\cite{alveolink} and Coyote~\cite{coyote} feature HLS-based RDMA offloading engines, as indicated in Table~\ref{tab:networking_platforms}. AlveoLink's RDMA engine~\cite{alveolink}, with AXI4-Streaming interfaces, lacks comprehensive information regarding its supported RDMA operations. In contrast, Coyote~\cite{coyote} designed a simplified HLS-based version of RoCEv2 RDMA, supporting RDMA read/write/send operations.

Compared to existing FPGA-based networking platforms~\cite{alveolink},\cite{VNx},\cite{copa},\cite{fpganic},\cite{coyote}, RecoNIC is an FPGA-based SmartNIC platform that can utilize a software network stack to manage non-RDMA traffic across various transport-layer protocols without requiring hardware offloading engines for all of them. When considering state-of-the-art FPGA-based SmartNIC platforms~\cite{azure_networking},\cite{open_nic},\cite{corundum},\cite{qep_driver}, RecoNIC provides an adaptive SmartNIC infrastructure, offering both a RoCEv2 RDMA offloading engine as well as lookaside and streaming compute blocks for network-attached acceleration.

\section{RecoNIC Platform}
\label{sec_reconic_framework}

\begin{figure}[t!]
	\centering
	\includegraphics[width=\columnwidth]{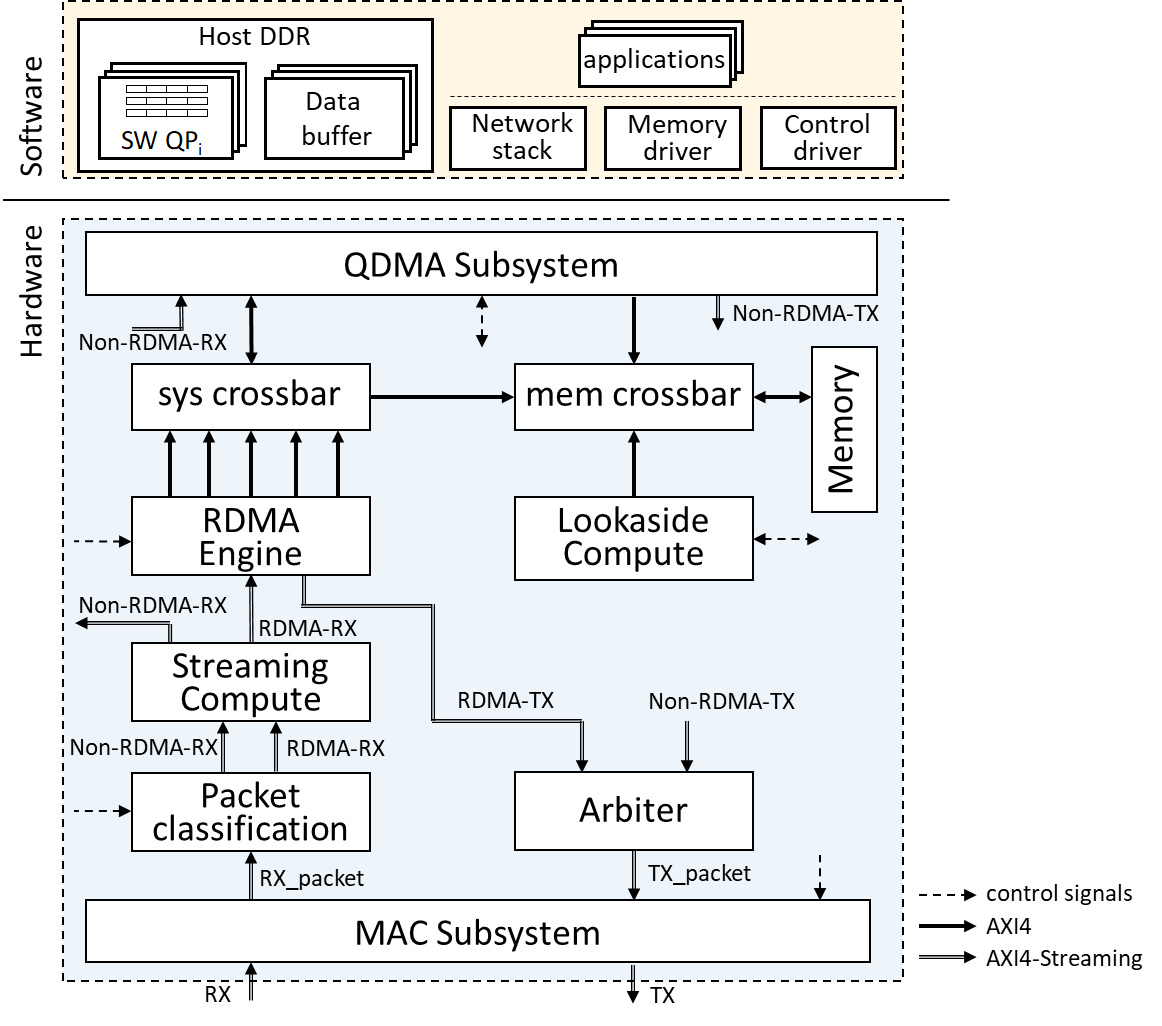}
	\caption{RecoNIC platform.}
	\label{reconic_arch}
\end{figure}

The RecoNIC platform is depicted in Fig.~\ref{reconic_arch} and consists of both hardware and software elements. The hardware shell includes a packet classification module, an RDMA engine, two programmable compute blocks (Lookaside and Streaming Compute) and a basic NIC module including a MAC subsystem and a DMA subsystem (QDMA). Additionally, it contains glue logic such as system/memory crossbars and an arbiter.

The RDMA engine is responsible for processing RDMA traffic, providing the capability to access data in either host or device memory of a remote peer connected via the network. User-defined accelerators implemented in the Streaming Compute and Lookaside Compute blocks can directly process data, including network-received data, within the device memory.

On the software side, there are three integral components: (1) network driver, which manages non-RDMA traffic like TCP/IP, UDP/IP, QUIC, and ARP; (2) memory driver, which facilitates seamless memory transfers between the host and device memories; and (3) control driver, which is responsible for configuring and controlling various hardware shell components.

\subsection{RDMA Engine}
\label{sec_rdma_engine}
RecoNIC leverages AMD ERNIC IP~\cite{ernic} as its RDMA offloading engine, which is designed in compliance with the RoCEv2 standard. ERNIC IP originally serves as an embedded RDMA soft IP with control operations orchestrated by either ARM or MicroBlaze processor. The distinctive feature of RecoNIC is that it interfaces ERNIC IP with x86 host CPUs, in addition to programmable compute blocks. The RDMA engine has two AXI4-Streaming input/output interfaces for RDMA RX/TX packets and five AXI4-MM interfaces for retrieving work queue elements, handling payload data, and recording completion queue entries. Furthermore, configuration of the RDMA engine is done through the PCIe AXI4-Lite interface.

Another distinctive feature of RecoNIC is the flexibility to allocate the queue pairs (QPs) and payload buffers in either host memory or device memory. The RDMA engine can access these QPs and buffers through the {\it sys\_crossbar} and {\it mem\_crossbar} based on the provided address offset. For example, in the current implementation, the address range of the device memory (16GB DDR4) spans from 0xa350000000000000 to 0xa3500003ffffffff, resulting in a 12-bit MSB mask (i.e., 0xa35). Thus, any QP or payload buffer address with 0xa35 MSBs will be read from the device memory.

\subsection{Programmable Compute Blocks}
\label{sec_comp_boxes}
There are two types of compute blocks: the {\it lookaside compute (LC)} block and the {\it streaming compute (SC)} block. The LC block is tailored for accelerators that need memory access, such as matrix multiplication. In contrast, the SC block is designed for applications that need to process incoming packets in a streaming mode, such as packet processing and network telemetry.

\subsubsection{Lookaside Compute (LC)}
\label{sec_lookaside}
\begin{figure}[t!]
	\centering
	\includegraphics[width=\columnwidth]{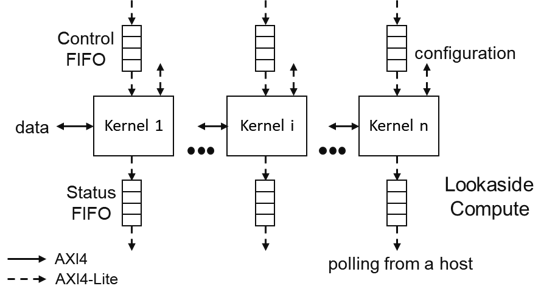}
	\caption{Lookaside compute block.}
	\label{lookaside_block}
\end{figure}

The LC block, as illustrated in Fig.~\ref{lookaside_block}, has the capacity to accommodate multiple kernels. These kernels can be crafted using either HLS tools or RTL coding. Each kernel is equipped with a control FIFO and a status FIFO. The control FIFO receives control messages from the host CPU. A control message is similar to an argument list when invoking a C function. For example, it can be a data structure consisting of a unique workload ID, the number of address arguments, and those addresses as arguments. To initiate a kernel, the host CPU can send a control message to the control FIFO via the AXI4-Lite interface. Once the control FIFO is not empty, the kernel retrieves a control message and begins execution. Using addresses from the control message, the kernel can access data stored in the host or device memory through the AXI4 data interface. Kernels can support multiple AXI4 interfaces to enhance memory access bandwidth and can have memory-mapped registers accessible to the host CPU via the AXI4-Lite interface.

Upon completing execution, a kernel signals its status via the associated status FIFO. The empty signal from a status FIFO can be connected to either the PCIe/host's interrupt system or a polling system using memory-mapped registers. In a design with an interrupt system, the LC kernel triggers the system's interrupt when the status FIFO is not empty, leaving the host to handle the interrupt accordingly. In a polling system design, the LC kernel writes a completion signal to a dedicated memory-mapped register when its output is ready in the status FIFO. The host monitors the dedicated register's value and responds accordingly. See section~\ref{sec_mm} for an illustration of how to design kernels within the LC block.


\begin{figure}[t!]
	\centering
	\includegraphics[width=0.5\columnwidth]{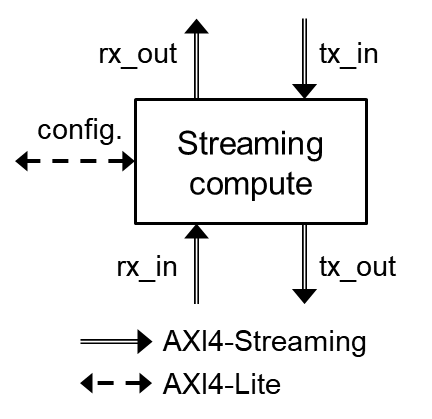}
	\caption{Streaming compute block.}
	\label{streaming_block}
\end{figure}

\subsubsection{Streaming Compute (SC)}
\label{sec_streaming}
The SC block, shown in Fig.~\ref{streaming_block}, is designed for applications that may need to process network traffic in a streaming mode. It has AXI4-Streaming interfaces for ingress and egress network traffic. Kernels in SC block can be implemented using Vitis Networking P4~\cite{vitisnetp4}, HLS or RTL coding. See Section~\ref{sec_pc} for an illustration of how to design kernels within the SC block.

\subsection{Packet Classification}
\label{sec_packet_class}
The {\it Packet Classification} module is designed to classify incoming network traffic into RDMA and non-RDMA traffic. RDMA traffic is handled by the RDMA engine, while non-RDMA traffic is redirected to the QDMA subsystem to be forwarded to and processed by the network driver on the host.

\subsection{Software Stack}
\label{sec_sw_stacks}
\begin{figure}[t!]
	\centering
	\includegraphics[width=\columnwidth]{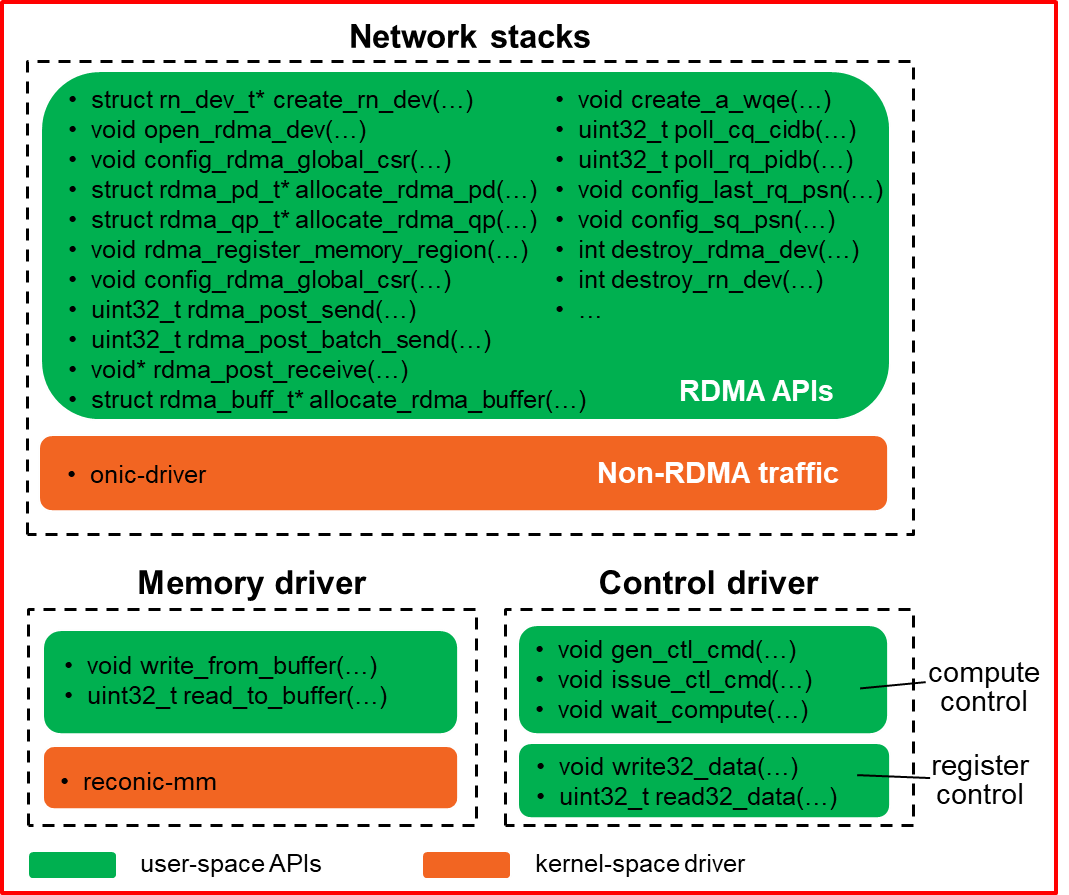}
	\caption{RecoNIC software stack.}
	\label{reconic_sw}
\end{figure}

RecoNIC's software stack, as depicted in Fig.~\ref{reconic_sw}, comprises of kernel-space drivers and user-space APIs.

{\it 1) Kernel-space drivers} consist of {\it onic-driver} and {\it reconic-mm}. The {\it onic-driver}~\cite{onic_driver} is a network driver that is based on AMD qep-driver~\cite{qep_driver}, OpenNIC driver~\cite{open_nic}, and libqdma~\cite{libqdma} to support non-RDMA traffic. The {\it reconic-mm} is a character device driver used to process device memory read/write requests from the host. The {\it onic-driver} is extended in RecoNIC with QDMA AXI4-Memory-Mapped channel support and works with {\it reconic-mm} to enable device memory access.

{\it 2) User-space APIs} comprise of {\it Memory API}, {\it Control API} and {\it RDMA API}. The {\it Memory API} is used to access device memory from the host, while the {\it compute control} and {\it register control} APIs in the {\it Control API} are designed for Lookaside Compute block's control and register configuration, respectively. The {\it RDMA API} provides the necessary functionality to utilize the RDMA engine for sending and receiving RDMA traffic. The user-space APIs are compiled and grouped as the {\it libreconic} library. For more information on user-space APIs, refer to the doxygen documentation generated in the {\it lib} folder of RecoNIC repository.

\section{Examples}
\label{sec_builtin}
In this section, we describe several examples that are included in RecoNIC's repository: DMA test, RDMA test, networked systolic-array matrix multiplication and packet classification.

\subsection{DMA Test}
\label{dma_test}
The DMA test example evaluates the functionality of copying data between the host and device memory. It supports both read from and write to the NIC's memory while the host acts as a master. Its usage is shown in Listing~\ref{lst:dma_test}.

\begin{figure}[b]
\begin{lstlisting}[language=TeX, caption=DMA Test Example, label={lst:dma_test}]
> cd examples/dma_test
> make
> ./dma_test -help
usage: ./dma_test [OPTIONS]

  -d (--device) device
  -a (--address) the start address on the AXI bus
  -s (--size) size of a single transfer in bytes, default 32,
  -o (--offset) page offset of transfer
  -c (--count) number of transfers, default 1
  -f (--data infile) filename to read the data from (ignored for read scenario)
  -w (--data outfile) filename to write the data of the transfers
  -h (--help) print usage help and exit
  -v (--verbose) verbose output
  -r (--read) use read scenario (write scenario without this flag)
\end{lstlisting}
\end{figure}

\subsection{RDMA Test}
\label{rdma_test}
The RDMA test contains RDMA read, write, send/receive and batch read/write examples using libreconic APIs. These examples follow the client-server model, and the description below uses work queue element (WQE), send queue (SQ), receive queue (RQ), completion queue (CQ) and queue pair (QP; consisting of SQ, RQ and CQ) terminology.

\begin{itemize}
	\item {\it Read}: The client node issues an RDMA read request (one WQE per SQ doorbell ringing) to the server node. The server node then replies with the RDMA read response packet.
	\item {\it Write}: The client node issues an RDMA write request (one WQE per SQ doorbell ringing) to the server node directly.
	\item {\it Batch Read}: The client node issues a burst of RDMA read requests (a set of WQEs per QP with the same payload size) to the server node by ringing the corresponding SQ doorbell with the number of requests in the burst (batch size). The server node then replies with RDMA read response packets one after the other.
	\item {\it Batch Write}: The client node issues a burst of RDMA write requests (a set of WQEs per QP with the same payload size) to the server node directly by ringing the corresponding SQ doorbell with the batch size.
	\item {\it Send/Receive}: The server node posts an RDMA receive request, waiting for an RDMA send request to its allocated RQ. The client node then issues an RDMA send request to the server node.
\end{itemize}

Listing~\ref{lst:rdma_read} shows the usage of these scenarios. A QP can be allocated at either host memory or device memory by indicating '-l host\_mem' or '-l dev\_mem'.

\begin{figure}[t]
\begin{lstlisting}[language=TeX, caption=Usage of an RDMA Read Example, label={lst:rdma_read}]
> ./read -h
usage: ./read [OPTIONS]

  -d (--device) character device name (defaults to /dev/reconic-mm)
  -p (--pcie_resource) PCIe resource
  -r (--src_ip) Source IP address
  -i (--dst_ip) Destination IP address
  -u (--udp_sport) UDP source port
  -t (--tcp_sport) TCP source port
  -q (--dst_qp) Destination QP number
  -z (--payload_size) Payload size in bytes
  -b (--batch_size) Batch size, number of WQEs per QP
  -l (--qp_location) QP/mem-registered buffers' location: [host_mem | dev_mem]
  -s (--server) Server node
  -c (--client) Client node
  -g (--debug) Debug mode
  -h (--help) print usage help and exit 
\end{lstlisting}
\end{figure}

\subsection{Lookaside Compute: Matrix Multiplication}
\label{sec_mm}
The lookaside compute example depicts the working of a network-attached matrix multiplication (MM) kernel. The MM computation is based on a systolic-array implementation in HLS C (following one of the examples in Vitis~\cite{vitis_accel_example}). The current implementation uses two peers where the data is stored on peer1, while the MM computation is done on peer2 which is equipped with RecoNIC (acting as the SmartNIC).

\begin{figure}[t]
	\centering
	\includegraphics[width=\columnwidth]{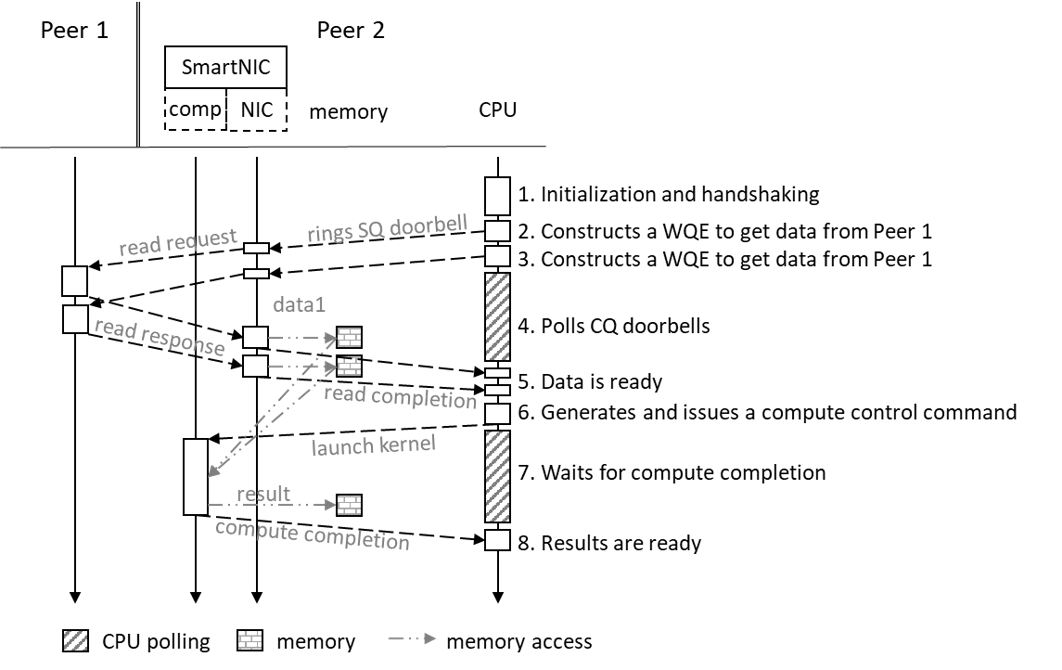}
	\caption{The workflow of a lookaside compute example: networked systolic-array matrix multiplication.}
	\label{lookaside_workflow}
\end{figure}

Fig.~\ref{lookaside_workflow} illustrates the workflow. Peer2 fetches data from peer1, performs the computation, and then notifies its host CPU. At step~\circled{1}, the host CPU initializes the system, sets up the connection, and exchanges information with peer1. In steps~\circled{2} and \circled{3}, the CPU constructs WQEs in an allocated SQ and rings the corresponding SQ doorbell which will trigger the RDMA engine in RecoNIC to send the read requests. The CPU then waits for data from peer1 by polling the corresponding CQ doorbells in step~\circled{4}. Once the RDMA engine in RecoNIC stores the data in its memory, it notifies the host CPU by issuing read completion signals on the respective CQ doorbells in step~\circled{5}. Then, in step~\circled{6}, the host CPU generates a compute control command for the kernel in lookaside compute block. During the kernel execution, the CPU waits for compute completion signal via polling or interrupt in step~\circled{7}. Once the computation is done and results are stored in memory, in step~\circled{8}, the CPU can then continue to handle next computation request.

In the current RecoNIC implementation, the host is responsible for all the control operations. However, these control operations can be offloaded to RecoNIC in future.

\subsection{Streaming Compute: Packet Classification}
\label{sec_pc}
The {\it packet classification} (PC) block in {\it shell/packet\_classification} directory illustrates an example of a streaming compute block, where incoming network packets are classified into RDMA and non-RDMA packets. Non-RDMA traffic is forwarded to the host (through QDMA AXI-Streaming interface), while RDMA traffic is handled by the RDMA engine. The PC block is written in P4 language to parse packet headers, and generate relevant metadata that can be used to classify the traffic. Some examples of packet header fields include Ethernet, IP, UDP, RoCEv2 base transport header (BTH), RDMA extended transport header (RETH), ACK extended transport header (AETH), immediate data extended transport header (ImmDt), and invalidate extended transport header (IETH). The P4 implementation is then converted into an RTL implementation with Vitis Networking P4 (VitisNetP4 for short, and formerly known as SDNet)~\cite{vitisnetp4} available in Vivado. The RTL implementation is then integrated into RecoNIC design. For more information, please refer to {\it shell/packet\_classification/packet\_parser.p4}.

\section{Hardware Simulation}
\begin{figure}[t]
	\centering
	\includegraphics[width=\columnwidth]{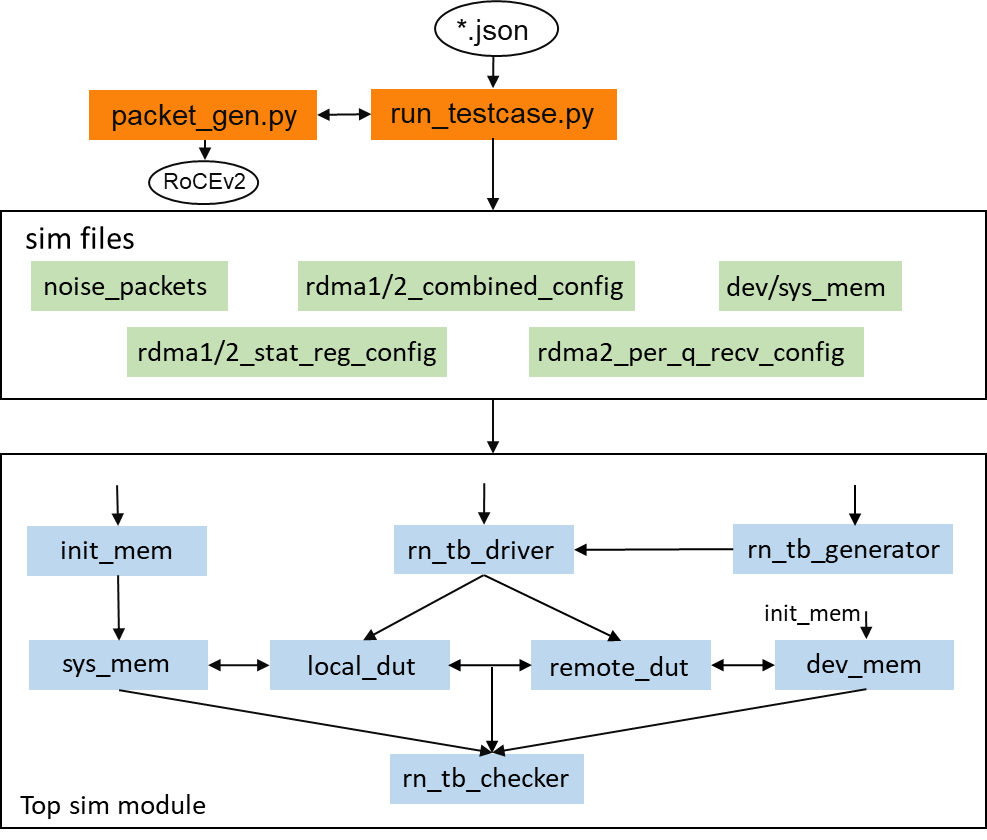}
	\caption{Hardware simulation framework.}
	\label{hw_sim}
\end{figure}

The hardware simulation framework is shown in Fig.~\ref{hw_sim}. The idea behind the framework is to leverage Python to generate RDMA configuration files based on a user-defined JSON file and let the hardware testbench modules configure the RDMA engine accordingly. Consequently, users can create multiple testcases by only changing JSON configuration files, resulting in a simpler and flexible framework for testing and debugging.

\begin{figure}[t]
	\begin{lstlisting}[language=TeX, caption=Usage of the Hardware Simulation Framework, label={lst:hw_sim}]
		> python run_testcase.py -h
		Usage:
		python run_testcase.py [options] regression,
		python run_testcase.py [options] -tc "testcase1 testcase2
		... testcasek"
		Options:
		-debug     : Debug mode
		-questasim : Use Questa Sim as the simulator. Default is 
		Vivado XSIM
		-roce      : Generate configuration files for RDMA 
		simulation
		-no_pktgen : Run testcases without re-generating packets
		-no_sim    : Only run analysis on the previous simulation 
		results
		-gui       : Use gui mode with the simulator
	\end{lstlisting}
\end{figure}

The simulation framework supports regression test through the {\it run\_testcase.py} script. The stimulus, control metadata and golden data are generated from {\it packet\_gen.py} script, where users can specify their own JSON file to generate a new testcase under {\it sim/testcases} directory. The {\it run\_testcase.py} script will automatically read those generated files and construct packets in AXI-streaming format along with all the control-related signals (e.g., RDMA configuration files include configuration for global control registers, memory registration, QP and WQE registers, and statistics registers). For cycle-accurate simulation, the framework supports both Questa simulator (questasim) and Vivado simulator (xsim). For more information, try a simulation example with the 'debug' option as shown in Listing~\ref{lst:hw_sim}, and check the generated files located at {\it sim/testcases/your\_sim\_example}.

\section{Experimental Evaluation}
\subsection{System Requirement}
All the experiments below are tested on machines with the following setup:
\begin{itemize}
	\item Two servers where each of them have an AMD Alveo U250 FPGA card connected via PCIe 3.0 x16 slot.
	\item The two Alveo U250 cards are connected via a 100Gb/s cable either directly or through a 100Gb/s switch.
	\item Ubuntu 20.04 and Linux kernel version 5.4.0-125-generic
	\item Python $>=$ 3.8
	\item Vivado 2021.2 with Vitis Networking P4 and ERNIC license
	\item Questa simulator 2021.3 (for simulation)
\end{itemize}

\begin{figure}[t]
	\centering
	\includegraphics[width=0.9\columnwidth]{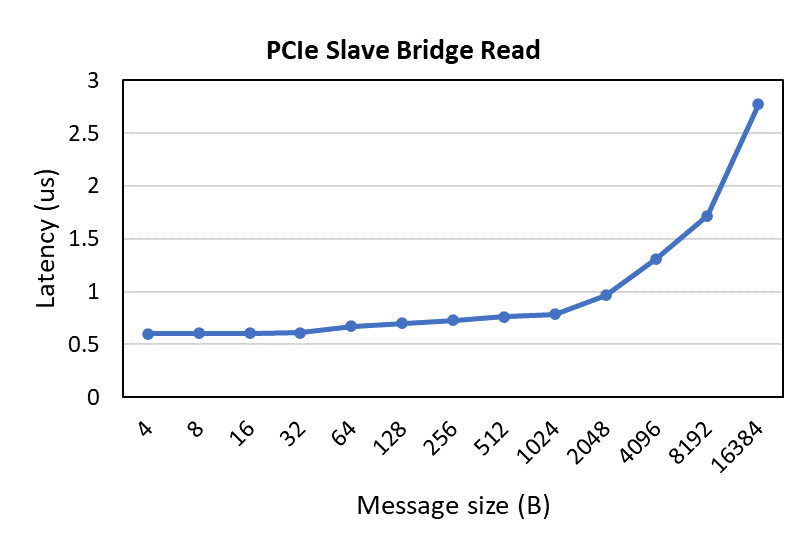}
	\caption{Latency for RecoNIC to host memory access.}
	\label{fpga_to_host_mem_lat}
	\vspace{-0.8em}
\end{figure}

\subsection{DMA Performance}
\subsubsection{Host as a master to access RecoNIC device memory} In this scenario, the host CPU accesses FPGA device memory over QDMA AXI4-MM channel. The effective throughput for DMA read and write is $\sim$13.00 GB/s and $\sim$13.07 GB/s respectively, which is 82.5\% of the theoretical peak throughput of PCIe 3.0 x16 slot.

\subsubsection{RecoNIC as a master to access host memory}
In this scenario, the FPGA logic acts as a master to access host memory over the QDMA slave bridge interface. Fig.~\ref{fpga_to_host_mem_lat} reports the latency as a function of message size, where FPGA takes $\sim$600 ns to $\sim$964 ns to access small messages ($<=$ 2048 B) from host memory.

\subsection{RDMA Read/Write Performance}
In RDMA read/write tests, we use two approaches: (1) measure performance per WQE, referred to as {\it single-request}; and (2) measure average performance for a batch of WQEs, referred to as {\it batch-requests}. In {\it batch-requests} approach, we create $n$ WQEs with the same payload size, ring the SQ doorbell once (for the last WQE) and poll CQ doorbell (for multiple completions), instead of ringing SQ doorbell and polling CQ doorbell for each WQE serially as done in {\it single-request}. We use $n=50$ and host CPU is responsible for preparation of WQEs, ringing of SQ doorbells and polling of CQ doorbells.

\begin{figure}[t]
	\centering
	\includegraphics[width=\columnwidth]{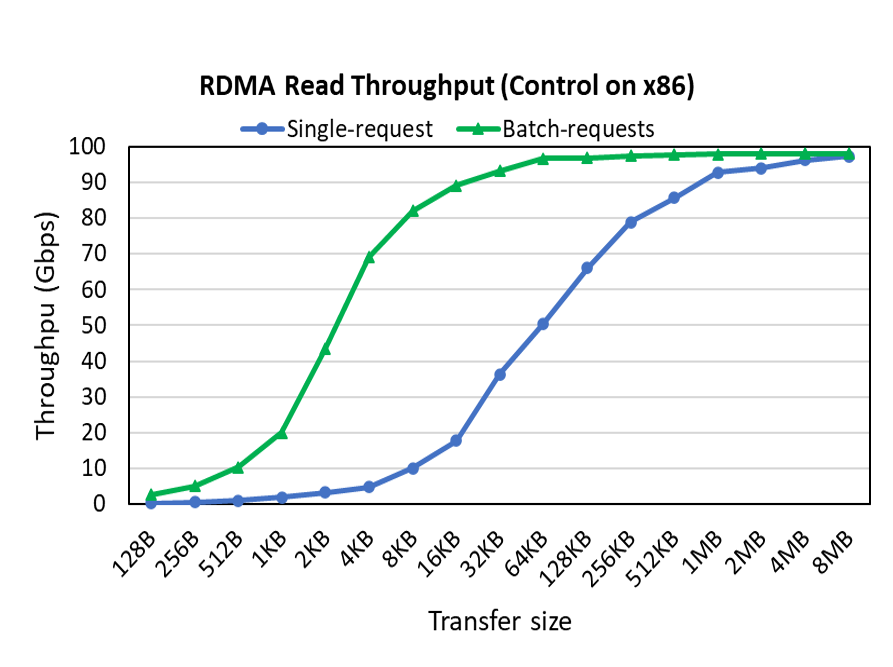}
	\caption{RDMA Read throughput with control on x86 CPU.}
	\label{read_throughput}
\vspace{-1.5ex}
\end{figure}

\begin{figure}[t]
	\centering
	\includegraphics[width=\columnwidth]{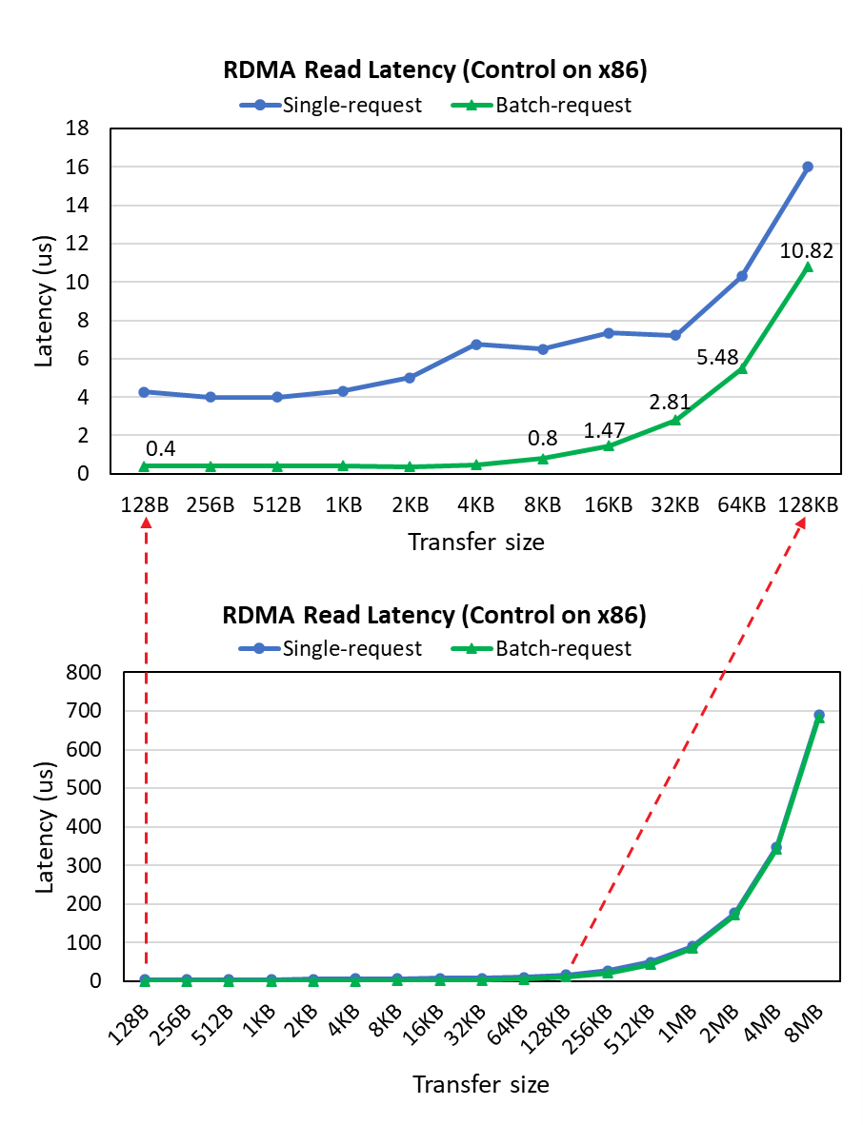}
	\caption{RDMA Read latency with control on x86 CPU.}
	\label{read_latency}
\vspace{-1.5ex}
\end{figure}

\begin{figure}[t]
	\centering
	\includegraphics[width=\columnwidth]{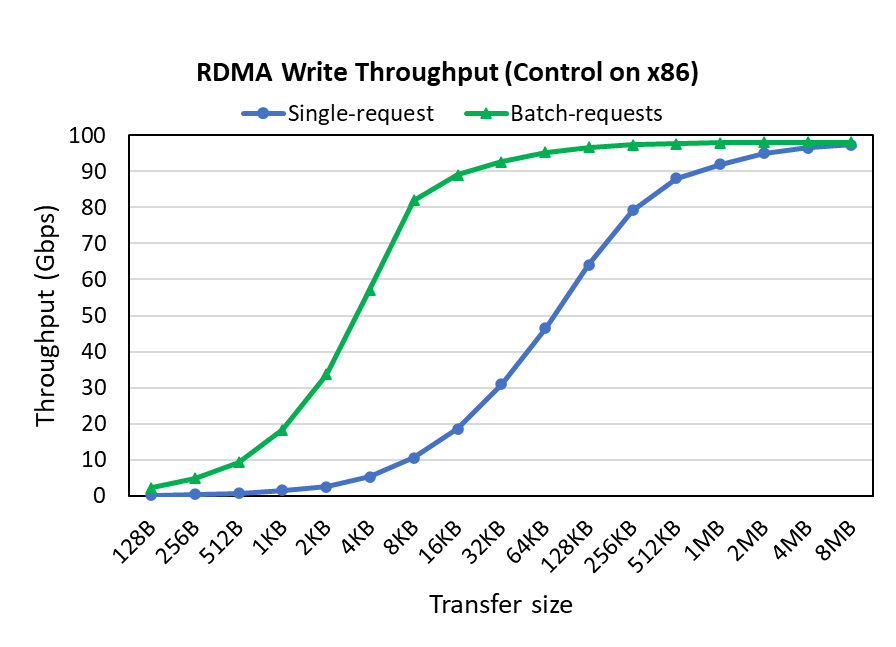}
	\caption{RDMA Write throughput with control on x86 CPU.}
	\label{write_throughput}
\vspace{-1ex}
\end{figure}

\begin{figure}[t]
	\centering
	\includegraphics[width=0.98\columnwidth]{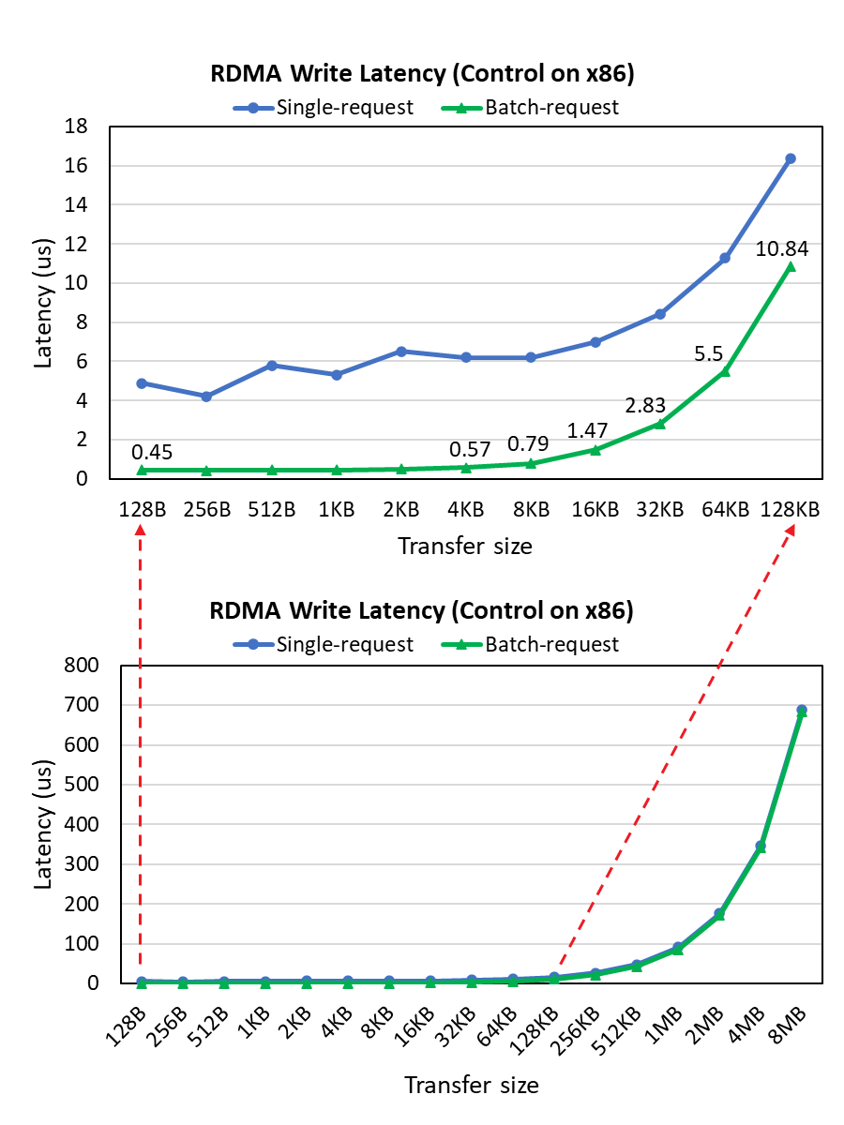}
	\caption{RDMA Write latency with control on x86 CPU.}
	\label{write_latency}
\vspace{-1ex}
\end{figure}

Fig.~\ref{read_throughput} and~\ref{read_latency} report RDMA read throughput and latency respectively as a function of data size transmitted. The throughput of {\it batch-requests} is much better than that of {\it single-request}, for example, when transferring 16 KB, {\it batch-requests} achieves $\sim$89 Gb/s compared to $\sim$18 Gb/s of {\it single-request}. Moreover, {\it batch-requests} can reach near line-rate (about 92 Gb/s) with much smaller data size of 32 KB. In terms of the latency, {\it batch-requests} can achieve almost 10x improvement compared to {\it single-request} when transmitting small data size ($<=$ 4 KB), which is $\sim$400 ns per RDMA read operation. The throughput and latency of RDMA write operation is shown in Fig.~\ref{write_throughput} and~\ref{write_latency} respectively, and the  trends similar to those of RDMA read operation.

The main reason for much better performance of {\it batch-requests} is that the overhead of SQ doorbell ringing and CQ doorbell polling is significantly reduced. Since doorbell ringing and polling is performed as register read/write operations over the PCIe AXI4-Lite interface, which are inherently slow, SQ doorbell ringing once and CQ doorbell polling once per $n$ WQEs results in significantly lower overall latency. Furthermore, when the RDMA engine notices that its SQ producer index doorbell is incremented by $n$, it issues $n$ read operations in a pipelined fashion. For example, although it takes the PCIe slave bridge $\sim$170 cycles (680 ns) to return the first WQE, the RDMA engine can receive subsequent WQEs every $\sim$10 cycles (40 ns). Thus, the overhead of long latency incurred by the PCIe bus can be amortized by issuing multiple WQEs together. A similar behavior is observed when writing multiple entries in CQ over PCIe bus.

\section{Conclusion}
This work introduced RecoNIC, which is a 100Gb/s FPGA-based SmartNIC platform powered with RDMA offloading engine and compute acceleration, and is available as open-source to the broader research community. We provided and explained built-in examples for the usage of RecoNIC, as well as discussed the hardware simulation framework for debugging and testing. Finally, we characterized the performance of RDMA read and write operations on RecoNIC. We hope that RecoNIC will enable researchers to build interesting network-attached acceleration use-cases.



%
%
%


\vspace{12pt}

\end{document}